 \documentclass[final,3p,times,twocolumn]{elsarticle}
\usepackage{amssymb}
\usepackage{lineno}
\usepackage{amsmath,amsfonts}
\usepackage{graphicx}
\usepackage{subfigure}
\usepackage[perc]{overpic} %scrivo sopra le immagini
\usepackage{gensymb}
\usepackage{xcolor,varwidth}

\journal{Nuclear Physics A}

\begin{document}

\begin{frontmatter}



\title{Recent studies on single-shot diagnostics for plasma accelerators at SPARC\_LAB}

\author[a]{F.G. Bisesto}
\author[a]{M.P. Anania}
\author[b]{M. Botton}
\author[a]{M. Castellano}
\author[a]{E. Chiadroni}
\author[c,d]{A. Cianchi}
\author[a]{A. Curcio}
\author[a]{M. Ferrario}
\author[e]{M. Galletti}
\author[b]{Z. Henis}
\author[a]{R. Pompili}
\author[b]{E. Schleifer}
\author[a]{V. Shpakov}
\author[a,b]{A. Zigler}

\address[a]{INFN Laboratori Nazionali di Frascati, Via Enrico Fermi 40, 00044 Frascati, Italy}
\address[b]{Racah Institute of Physics, Hebrew University, 91904 Jerusalem, Israel}
\address[c]{University of Rome "Tor Vergata", Physics Department, Via della Ricerca Scientifica 1, 00133 Roma, Italy}
\address[d]{INFN-Roma Tor Vergata, Via della Ricerca Scientifica 1, 00133 Roma, Italy}
\address[e]{Istituto Superior Tecnico de Lisboa, 1049-001 Lisbon, Portugal}

\begin{abstract}

Plasma wakefield acceleration is the most promising acceleration technique for compact and cheap accelerators, thanks to the high accelerating gradients achievable. Nevertheless, this approach still suffers of shot-to-shot instabilities, mostly related to experimental parameters fluctuations. Therefore, the use of single shot diagnostics is needed to properly understand the acceleration mechanism.
In this work, we present two diagnostics to probe electron beams from laser-plasma interactions, one relying on Electro Optical Sampling (EOS) for laser-solid matter interactions, the other one based on Optical Transition Radiation (OTR) for single shot measurements of the transverse emittance of plasma accelerated electron beams, both developed at the SPARC\_LAB Test Facility.
\end{abstract}


\end{frontmatter}

%\linenumbers

\section{Introduction}

The possibility to reach power densities larger than $10^{18}\ W/cm^2$ at femtosecond level, thanks to the important achievements in laser technology, has opened the way for future compact accelerators \cite{tajima1979laser,wilks2001energetic}. 
Even though compact accelerating systems have been already demonstrated \cite{geddes2004high,mangles2004monoenergetic,leemans2006gev,maksimchuk2000forward,macchi2013ion,clark2000energetic,snavely2000intense} and new x-ray radiation sources has been developed \cite{curcio2017first,rousse2004production,fuchs2010laser}, the produced charged particle beams are still affected by shot-by-shot instabilities. Moreover, the physical mechanism for ion acceleration, exploiting the interaction of high power lasers with solid matter, is not clear yet. 
Therefore, single-shot diagnostics play an important role in order to achieve a better control in laser-plasma experiments. 

In this work, we present two diagnostics tools, developed at SPARC\_LAB Test Facility \cite{ferrario2013sparc_lab}, suitable to investigate the properties of electrons produced by high intensity laser - matter interaction. One relying on Electro Optical Sampling (EOS) has been employed to probe the fast electron temporal profile, produced during the interaction between the high intensity laser FLAME and solid state matter.
%%
%EO effect has been already used as bunch length diagnostics for electron beams from laser wakefield acceleration \cite{jamison2003high,helle2012extending}, but it has been never employed to probe laser-solid target interactions. In particular, the large amount of electrons, released from the opposite surface with respect to the laser-target interaction, called \emph{fast electrons}, are responsible for the surface ionization and consequently ion and proton acceleration. They have been extensively studied in the past with different approaches \cite{popescu2005subfemtosecond,jung2005study,santos2002fast,santos2007fast,storm2009high,bellei2010micron}.
%Nevertheless, with this technique, we succeed to measure for the first time the temporal profile of the electric field carried by fast electrons generated during the interaction with an unprecedented resolution below $100$ fs \cite{pompili2016sub,bisesto2017innovative}. Furthermore, we have investigated the role of the target shape in the fast electron emission in order to optimize the ion and proton acceleration process. In particular, we found out that structured targets allow to get a boost in accelerating field \cite{pompili2016femtosecond,bisesto2017innovative2}.
%%%%
On the other hand, we have studied the possibility to exploit Optical Transition Radiation (OTR) for a single-shot transverse emittance measurement of plasma accelerated electron beams. In particular, it is possible to evaluate the rms emittance correlation term  by means of a microlens array \cite{bisesto2017innovative,bisesto2017innovative2,cianchi2016transverse}. 
%This technique is very promising for future applications in laser wakefield accelerators, allowing to properly tune the experimental parameters by measuring shot-by-shot the quality of the accelerated beam.

%We present in this work an overview of the developed diagnostics, together with experimental results. 
%Particular emphasis is given to the first temporal snapshot of the electric field carried by fast electrons escaping from the interaction between ultra intense laser and solid target. Preliminary tests of the single shot emittance measurements  done at the SPARC$\_$LAB photo-injector are also presented and discussed.

\section{Longitudinal profile measurement of fast electron electric field}
Thin foils irradiated by high-intensity short-pulse lasers produce a large amount of particles, e.g. ions and protons with energies in multi-MeV range \cite{clark2000energetic,snavely2000intense,mackinnon2002enhancement}.
A key role in the acceleration process is played by the most energetic electrons, amoung those directly accelerated by the laser and passing through the target, who escape from its rear surface. Indeed, these particles, called \emph{fast electrons}, leave an electrostatic potential onto the target, due to the unbalanced positive charge left on it \cite{poye2015physics}. In turn, such potential generates an electric field that ionizes and accelerates surface ions in a process called \emph{Target Normal Sheath Acceleration} (TNSA) \cite{wilks2001energetic}.

\begin{figure}[htb!]
\centering
\includegraphics[width=0.9\columnwidth]{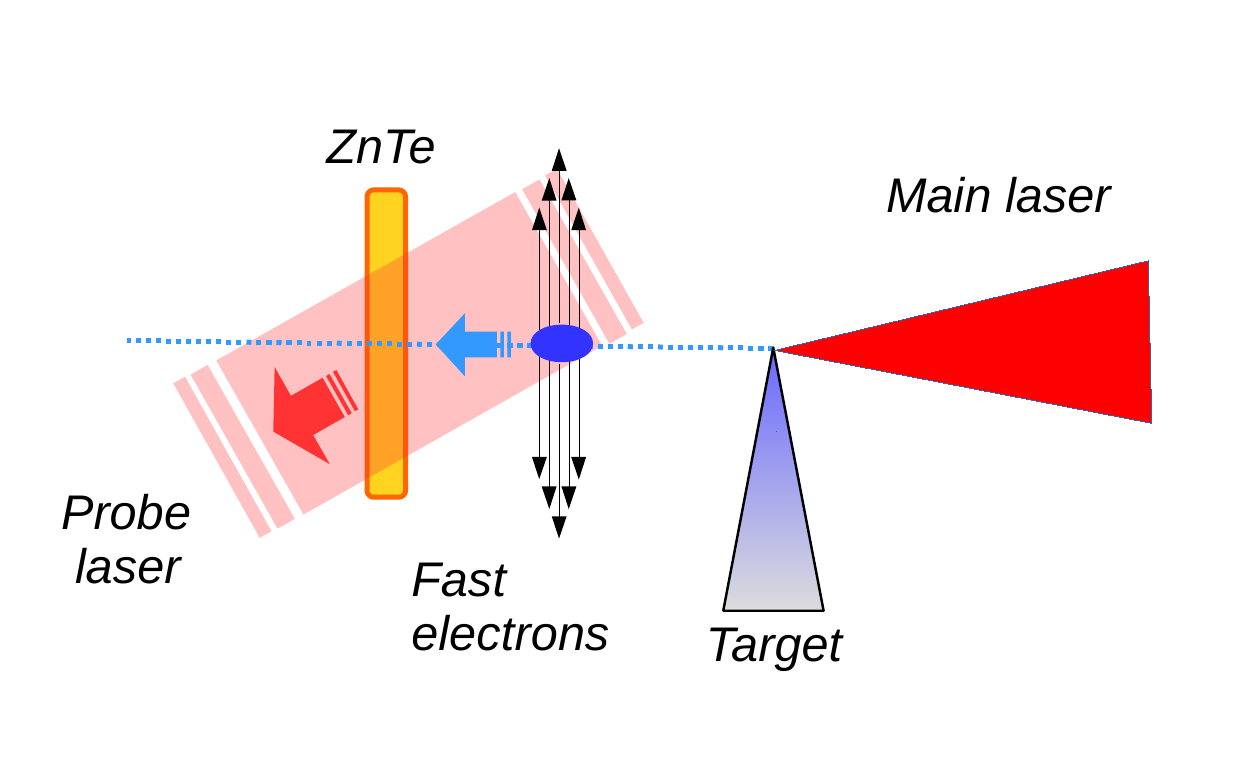}
\caption{Setup of our experiment. The FLAME laser is focused on the tip of a metallic target. The EOS diagnostics, based on a $500\ \mu m$ thick ZnTe crystal, placed $1$ mm downstream the target, allows to measure the temporal profile of the emitted electrons. For this purpose, a secondary laser beam, directly splitted from the main one, acts as a probe for the local birefringence induced by the fast electron electric field.}
\label{FlameSetup}
\end{figure}
In this work, we report about our diagnostics for temporal measurement of the electric field carried by fast electrons, exploiting a diagnostics based on Electro-Optical Sampling (EOS) \cite{wilke2002single,pompili2017electro,steffen2009electro}.
The experiment has been performed with the FLAME laser \cite{flame} at the SPARC\_LAB test-facility \cite{ferrario2013sparc_lab}. This system can deliver ultrashort pulses down to $25$ fs (FWHM), with energy up to $5$ J on target at $800$ nm central wavelength and at $10$ Hz repetition rate. Thanks to a $f/10$ off-axis parabolic mirror with focal length $f=1$ m, it is possible to reach peak intensities of about $10^{20}W/cm^2$.
Moreover, an ancillary laser line is directly split from the main laser and it is employed as test beam in pump-probe experiments. The synchronization between the two lasers is achieved by performing an optical autocorrelation: this represents our reference time.

%The probe laser employed to detect the EO signal ($35$ fs pulse duration) is directly split from the FLAME laser, ensuring a jitter-free synchronization.  The temporal synchronization with the main, i.e. our reference time, is achieved through autocorrelation. 
Figure \ref{FlameSetup} shows the experimental set-up, based on EOS spatial decoding technique \cite{cavalieri2005clocking}, with the probe laser entering into the crystal at an incidence angle $\theta_i\approx 28^{\circ}$. In this way, the temporal charge profile of the emitted electrons is spatially imprinted along the transverse profile of the probe laser \cite{pompili2016sub}. Being $d_L\approx 6$ mm (FWHM) its transverse spot size, the effective time window is $\Delta t = (d_L/c) \cdot \sin\theta_i\approx 10$ ps, where $c$ is the speed of light in vacuum, with a resolution of about $100$ fs.

\begin{figure}[htb!]
\centering
\includegraphics[width=0.9\columnwidth]{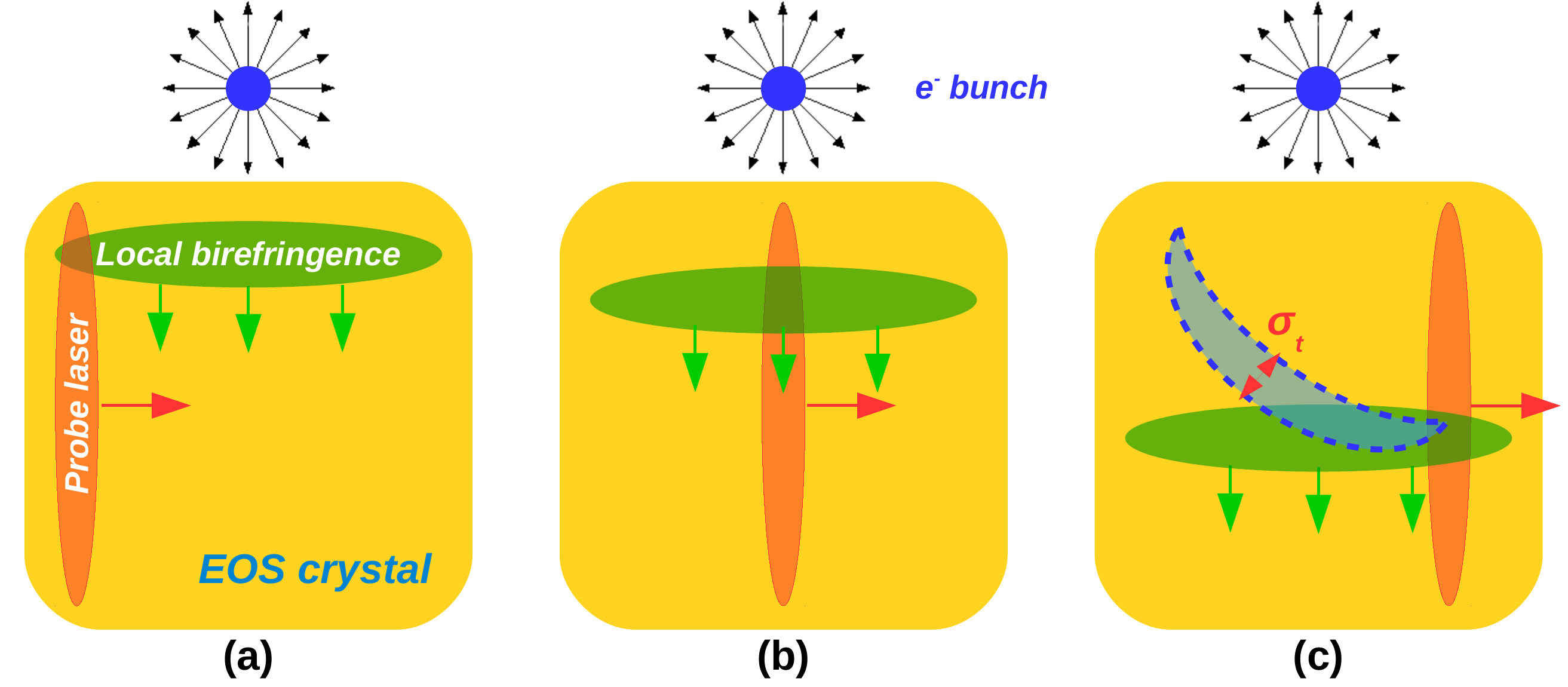}
\caption{Spatial encoding process: (\textbf{a}) the electric field carried by the electrons induces a birefrigence into the crystal; (\textbf{b}) while it propagates inside the crystal, the local birefringence shifts downwards; (\textbf{c}) the probe laser crosses the crystal and its polarization is consequently rotated. The measured signal comes from the area where the local birefringence and the probe laser are temporally overlapped (blue region) \cite{pompili2016sub}.}
\label{fig:eosencoding}
\end{figure}
\begin{figure}[h]
\centering
\includegraphics[width=0.9\columnwidth]{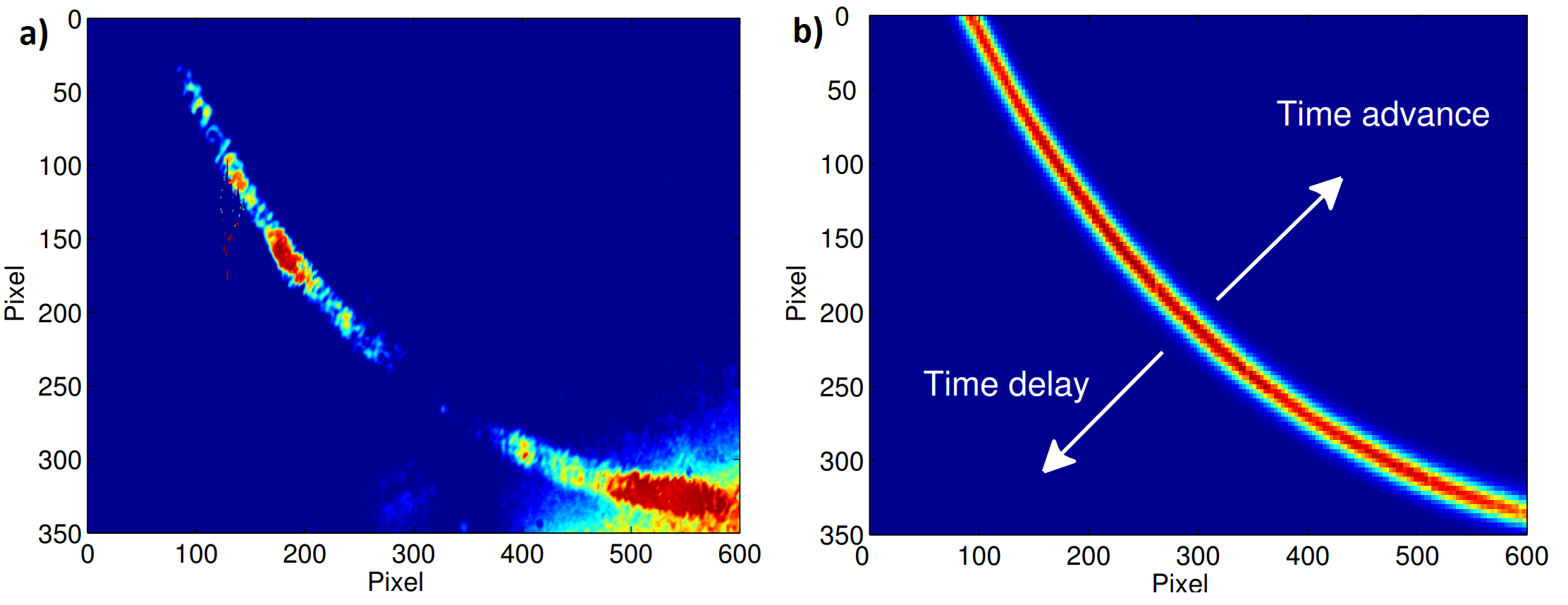}
\caption{a) Typical measured and b) simulated EOS signals. The not perfect uniformity in the experimental signals is mainly due to inhomogeneities present both on the ZnTe crystal surface and on the transverse profile of the probe laser.}
\label{fig:eosresults}
\end{figure}
\begin{figure}[h]
\centering
\includegraphics[width=0.9\columnwidth]{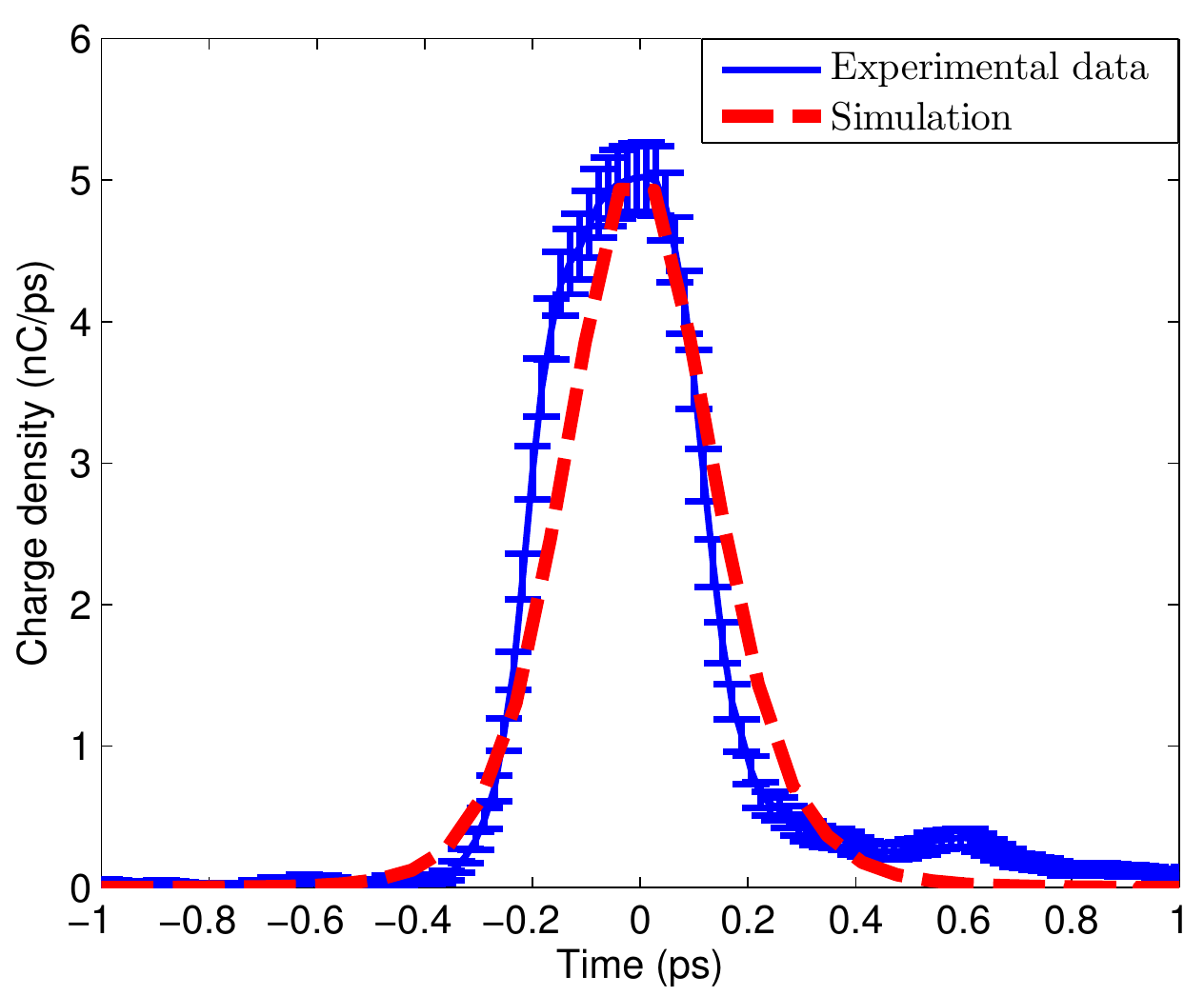}
\caption{Typical measured temporal profile. The profile is obtained by averaging different line-outs performed along the time direction of the Fig. \ref{fig:eosresults}a. The experimental data (blue line) are compared with the corresponding profile provided by the numerical simulation (red dashed line). The errorbars are calculated as the standard deviation of line-outs average. \cite{pompili2016sub}}
\label{ChargeProfile}
\end{figure}

Figure \ref{fig:eosresults} shows some typical results, compared with simulations.
Due to our set-up geometry, with the electrons moving normally to the ZnTe crystal and the probe laser propagating laterally from right to left, as in Fig. \ref{fig:eosencoding}, the expected signal is curved. This has been confirmed by simulations and experiment (Fig. \ref{fig:eosresults}).% \cite{pompili2016sub}.
By measuring the relative delay with respect to the reference time, this diagnostics can also estimate the energy of the emitted electrons \cite{pompili2016sub}, providing measurements resolved in time.
Moreover, since the electric field, inducing the electro-optic effect, is proportional to the bunch charge, this can be estimated from the signal intensity \cite{pompili2017electro}, properly calibrated with respect to the probe laser energy.
Once the electron mean energy and charge are determined, the width of the measured signal is proportional to the bunch duration. Figure \ref{ChargeProfile} shows a typical longitudinal profile measurement. Here, it is reported the average of different line-outs performed along the time direction, parallel to the temporal window diagonal. In detail, in this shot, an electron bunch with $2.1$ nC charge, $14$ MeV energy and about $500$ fs duration has been detected.%Therefore, its temporal profile can be obtained by performing a line-out along the time direction depicted in Fig. \ref{11shot_exp}.

\section{Diagnostics for single-shot emittance measurements}
High gradient plasma structures provide accelerators at mm-scale. Nevertheless, the accelerated electron beams are still characterized by a large energy spread, around $10\%$, and shot-to-shot instabilities. These issues prevent the use of conventional, multi-shot diagnostics for the emittance measurements, such as the quadrupole scan technique \cite{lohl2006measurements,mostacci2012chromatic}. 

Our idea relies on Optical Transition Radiation (OTR), emitted when a charged particle crosses the boundary between two media with different refractive indices \cite{ter1972high}. Indeed, its angular distribution in the far field is sensitive to the beam divergence. In particular, starting from the angular distribution for a single particle \cite{ter1972high} and considering the convolution with a Gaussian distribution in angle, we obtain a higher central minimum, directly correlated with the rms beam divergence $\sigma^{\prime}$ \cite{bisesto2017novel}.
Despite previous work aiming to exploit OTR to measure electron beam emittance \cite{feldman1990developments},
our method relies on the analysis of the OTR angular distribution in the focal plane of a microlens array to measure also the rms emittance correlation term. Indeed, this optical element, focusing different portions of the radiation in different points, directly correlates electron transverse momentum with the spatial position. The principle is similar to the pepper-pot emittance measurement \cite{zhang1996emittance}, but, at the same time, our technique is more flexible. Indeed, it does not need a thick mask to be put on the electron beam cutting its phase space and, moreover, there is not a dependence on the screen position, that in our case is fixed in the microlens focal plane. 
\begin{figure}[htb!]
\centering
\includegraphics[width=0.9\columnwidth]{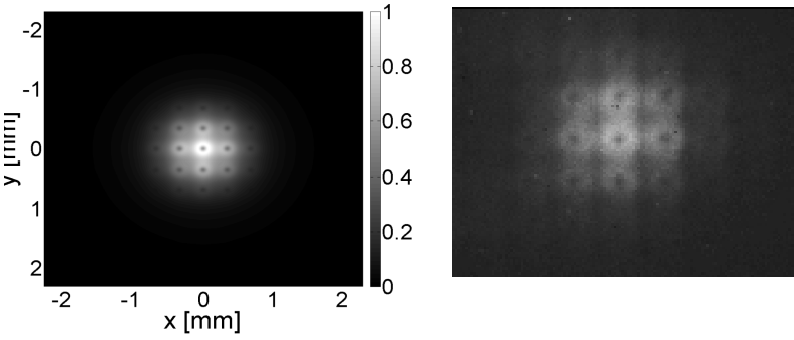}
\caption{Comparison between a Zemax simulation (left) and a typical experimental measurement (right).}
\label{fig:OTRsimulation}
\end{figure}

\begin{figure}[htb!]
\centering
\includegraphics[width=1\columnwidth]{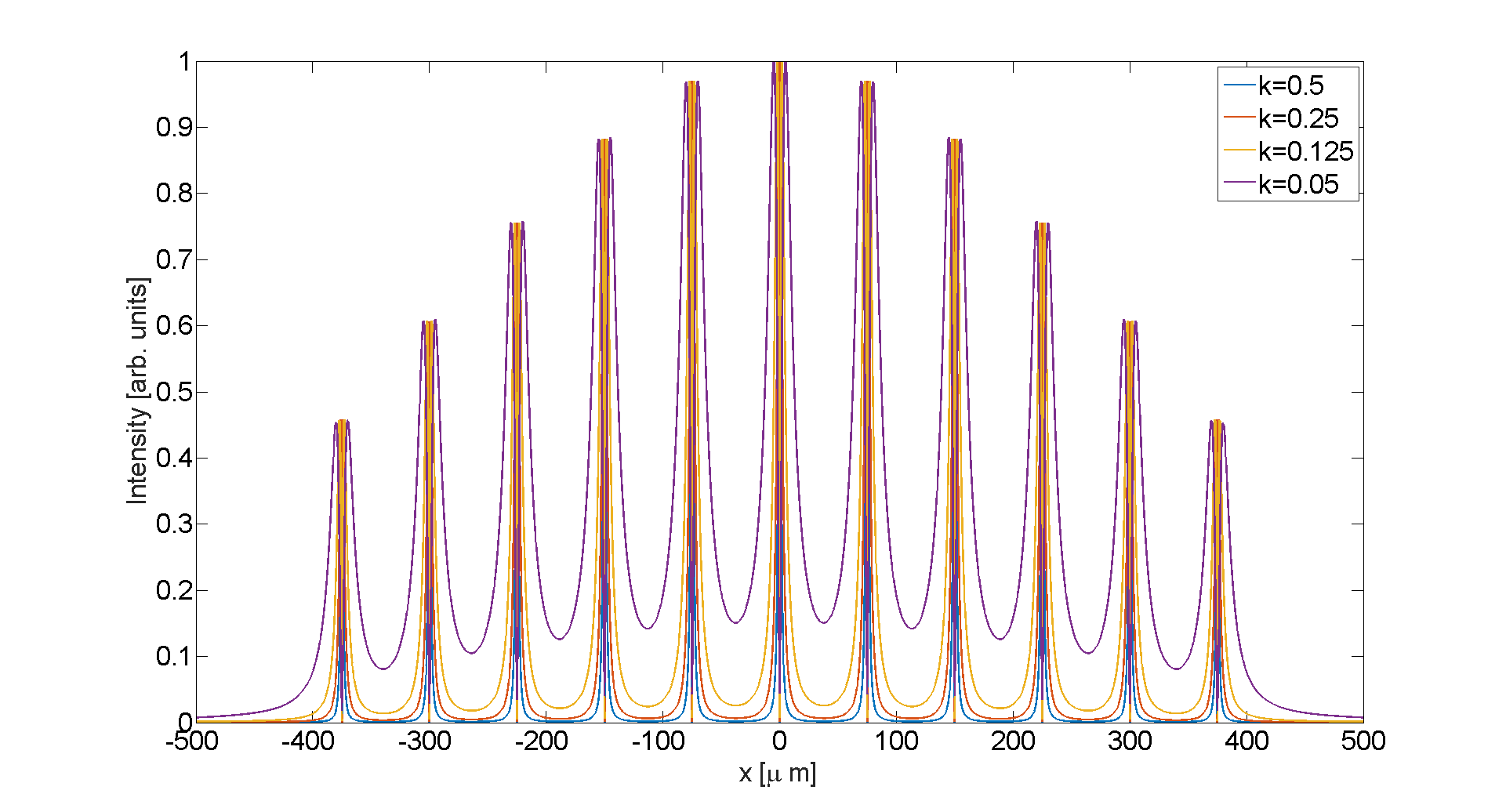}
\caption{Effect of the overlap between adiacent angular distributions in the microlens array focal plane. For this simulation, an electron beam with $\gamma=300$ has been considered.}
\label{fig:oseoverlap}
\end{figure}

Several simulations have been performed to prove the feasiblity of this technique, firstly to understand the behaviour of the OTR field with a microlens array. For this purpose, a custom code in Zemax has been created to simulate this radiation as produced from a whole electron bunch \cite{bisestozemax}. Simulation results are shown in Fig. \ref{fig:OTRsimulation}. The angular distributions coming from each microlens are clearly distinguishable.
Moreover, a design analysis of the microlens array to be used has been performed. In particular, effects due to the diffraction with microlens aperture and to the overlap between adiacent angular distribution have been considered.
Concerning the first one, microlenses with a diameter $d\ge \gamma\lambda$ are needed to avoid diffraction, while the overlap between their different focal planes is minimized when $k=d/f>>1/\gamma$, i.e. when the lobes of two different angular distributions are far enough (see Fig. \ref{fig:oseoverlap}).
The microlens array commercially available at that time satisfied the first requirement, but not the second one. Indeed, the focal length was too long to avoid the overlap between different angular distributions.

\begin{figure}[htb!]
\centering
\includegraphics[width=0.9\columnwidth]{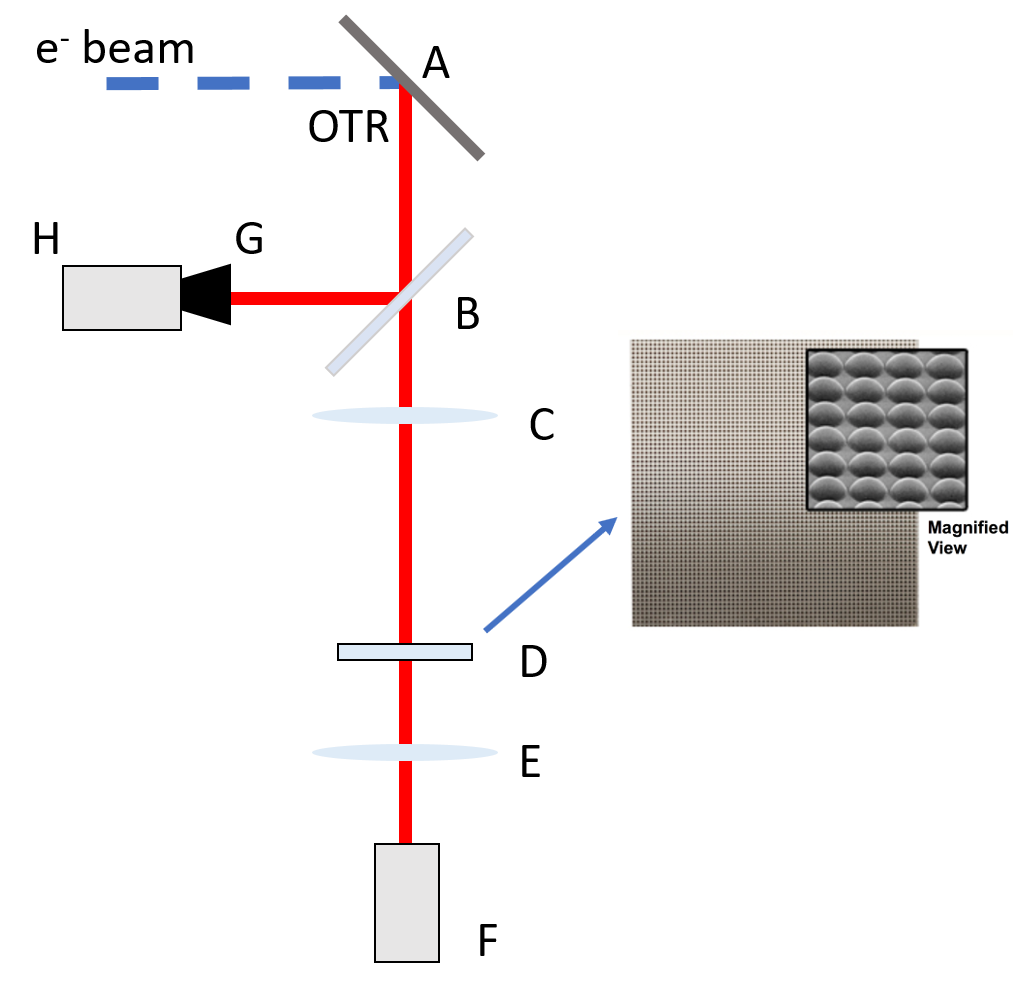}
\caption{The radiation produced by the electron beam impinging on an aluminum-coated %Pls define
silicon screen (\textbf{A}) can follow two different paths, thanks to a beam splitter (\textbf{B}). In one arm, the radiation is collected by a $f=$ $400$ mm achromatic doublet (\textbf{C}) whose image plane hosts a microlens array (\textbf{D}), $10$ mm $\times$ $10$ mm size with $300$ $\mu$m pitch, $18.7$ mm focal length plano-convex lenses. Microlens focal plane is imaged onto an intensified CCD camera (\textbf{F}) by means of a $50$ mm focal length achromatic doublet (\textbf{E}). The~second arm of the setup  is used to image the radiation produced on the metallic screen for transverse beam profile measurements on a high quantum efficiency CCD camera (\textbf{H}) by means of a Nikon $f=$ $180$ mm focal length F/$2.8$ objective (\textbf{G}).}
\label{fig:OTRsetup}
\end{figure}

A preliminary experiment has been performed with SPARC$\_$LAB photoinjector \cite{ferrario2013sparc_lab}, using $200\ pC$ bunch charge at $125\  MeV$. Figure \ref{fig:OTRsetup} shows the experimental set-up layout: the OTR coming from an aluminium-coated silicon screen (A) is divided by means of a beam splitter (B); an optical line hosts a high quantum efficiency CCD camera (G) to image the source transverse profile, while a $400$ mm achromatic doublet (C) in the second arm transfers the image of source on a microlens array (D); a second $50$ mm achromatic doublet (E) is used to image the microlens focal plane onto an intensified CCD camera (F).
Although we could not measure the electron bunch emittance, since its energy has limited our angular resolution \cite{cianchi2016transverse}, these preliminary results (Fig. \ref{fig:OTRsimulation}) have shown that it is possible to produce the OTR angular distribution from different portions of the beam, as expected from our simulations.

\section{Conclusions}
We presented two single-shot diagnostics for electron beams generated from ultra intense laser-plasma interactions developed at SPARC\_LAB Test Facility \cite{ferrario2013sparc_lab}.
One, relying on Electro Optical Sampling (EOS), has allowed to measure the longitudinal profile of electric field carried by fast electrons generated during the interaction of a high intensity laser with solid matter. The characterization of these electrons can help to better understand the process of ion acceleration occurring in this kind of interaction. 
As reported in Fig. \ref{fig:eosresults}, the signals have been well reproduced and electron properties, in terms of charge, current and mean energy, have been retrieved. Besides the first measurement of longitudinal profile of the fast electrons, shown in Fig. \ref{ChargeProfile}, this diagnostics has been exploited to study the influence of geometrical shape of the target. In particular, we have shown that faster electrons and with more charge, resulting in a stronger electric field, are emitted from structured target \cite{pompili2016femtosecond}. Therefore, it is possible to boost the ion acceleration by using structured targets exploiting the consequently electric field enhancement.

On the other hand, the properties of Optical Transition Radiation (OTR) have been studied to design a single-shot transverse emittance measurement. In particular, our idea relies on using a microlens array, analyzing the OTR angular distribution coming from each microlens, placed in a specified position. In this way, by measuring also the transverse spot size (see Fig. \ref{fig:OTRsetup}), an emittance measurement is achievable in one shot. This kind of diagnostics represents a very useful tool to fully characterize an electron beam from LWFA despite its typical shot-to-shot instabilities and relative large energy spread.  Even though some preliminary tests, obtained with the SPARC\_LAB photo-injector, did not allow to measure the emittance because of poor angular resolution (due to a low electron energy), they have shown the possibility to measure the OTR angular distribution produced by different portions of the electron beam, as foreseen by our simulations.
In the next future, a measurement on a plasma-accelerated electron beam at FLAME Facility is still under design, with both self-injection and external injection \cite{bisesto2016laser,rossi2016stability} scheme, as well as a new experimental campaing at the SPARC\_LAB photoinjector, working at the nominal energy, will be performed. 
%The aim of this experiment, beside testing this diagnostics, is to quantify the emittance growth when the beam leaves the plasma channel by exploiting recent results with betatron radiation measurements \cite{curcio2017trace}.



\bibliographystyle{elsarticle-num} 
\bibliography{eaac17_bib}


%\begin{thebibliography}{00}


%\bibitem{}
%
%\end{thebibliography}
\end{document}